2022

# Systematic Mapping Protocol

AGILE STRATEGIES FOR SOFTWARE DEVELOPMENT ACCORDING TO TECHNICAL AND ENVIRONMENTAL COMPLEXITY FACTORS

FINAL VERSION: 24/09/2022

Fernando Pinciroli

Instituto de Investigaciones
Facultad de Ciencias Exactas, Físicas y Naturales
Universidad Nacional de San Juan, Argentina.

fernando.pinciroli@gmail.com

SYSTEMATIC MAPPING PROTOCOL
AGILE STRATEGIES FOR SOFTWARE DEVELOPMENT ACCORDING TO TECHNICAL AND ENVIRONMENTAL COMPLEXITY FACTORS

# Contents

# 1. Introduction

Software development projects management is a complex endeavor because it requires dealing with numerous unforeseen events that constantly arise along the way and that go against the expectations that had been established at the beginning. A good project leader is not so much who carries out what is planned, but rather who is able to deal with all the inconveniences that arise and, in the end, achieve a result that is as close as possible to what was expected [1].

In other words, what is most valued is the ability to adapt to changes, to face unforeseen events, to make the best decisions regarding a reality that is imposed as the project progresses.

On the other hand, those of us who dedicate ourselves to IT projects have the tendency to cling to the tools that gave us the best results, although many times they are not the most appropriate for the case or, if they are, we continue to maintain them even when the context of the project has changed.

Also, software development is complex because it is inherently so. According to Brooks, complexity is an essential, not an accidental, property of software [2]. It is also so because people, who are an essential and intensive part of their development, and our relationships, are even more so.

Finally, for some decades we have extrapolated many proven project management approaches, techniques, and tools from other fields to computer science [3]. Surely, many of them were of value, but others did not give the expected results, at least from the evidence of successes and failures in software development projects [4].

Even though many ideas, such as iterative development, existed before [5], the appearance of the Manifesto for Agile Software Development [6], in 2001, radically changed the way of seeing project management and was like a breath of fresh air for the discipline, but there are still several difficulties that, we believe, must be overcome:

- On many occasions, agile tools and techniques are used in projects, but they are not done with an agile approach.
- Frequently, reality is not analyzed to determine the best tools to be used; instead, project managers stick to the same recipe that previously gave better results or choose the tools before knowing the problem.
- Agility implies adaptation, but there are not too many cases in which tools and techniques are adapted or combined depending on the problem to be faced.
- Many people are trained in a single agile method and use it systematically, rather than taking an "agnostic" view.
- In many cases, what agility means is not understood, and it is confused with other ideas: life cycle models, methods, following a recipe, etc.

This is a non-exhaustive or prioritized list of difficulties, but we seek to describe the problem to which, through this study, we intend to characterize in order to look for solutions and improvements.

We are interested in portraying the state of the art of software development project management with respect to the selection of management strategies based on the complexity of the problem to be faced. Our goal is to collect all the available evidence, analyze it, and study the possibility of guiding in the selection of most appropriate approaches, techniques, and tools, depending on the complexity of the problem, to better manage projects.

Evidence-Based Software Engineering (EBSE) aims to convert the need for information into an answerable question, tracking down the best evidence with which to answer that question and critically appraising the evidence for its validity. Kitchenham et al. affirm that EBSE intends "to provide the means by which current best evidence from research can be integrated with practical experience and human values in the decision making process regarding the development and maintenance of software" [7]. In this document we detail the planning phase of a Systematic Mapping Study (SMS), used to structure the findings on a research area, based on the guidelines from Petersen et al. [8].

Our goal is twofold: to identify the evidence present in the scientific literature about criteria to select agile or plan-based project management approaches, techniques and tools, and the frameworks used to characterize the projects, in case they are used. We will perform a systematic mapping study of the literature for this purpose.

The rest of this article is structured as follows: in section 2 we describe the research method, section 3 presents the strategy to deal with validity threats, and, finally, section 4 offers our conclusions.

## 2. Research method

### 2.1. Goal and research questions

This work aims to identify and classify the to identify the criteria to select agile or plan-based project management approaches, techniques and tools, and the framework s used to characterize the projects, in case they are used.

It is of our particular interest to identify agile and plan-based approaches that could collaborate with this objective, regardless of whether they were already applied in the industry or if they are still in a study stage, without having reached their employment in real-settings.

Particularly we will consider:
- Project management approaches.
- Project management life cycle models.
- Characteristics to be considered in order to select a project management approach.
- Characteristics to be considered in order to select a project management life cycle model.
- The frameworks used to characterize project management approaches.

- Evidence of results and challenges present in the industry.

A set of Research questions (RQ) has been designed to accomplish this goal (see Table 1). Furthermore, a set of publication questions (PQ) has been included to characterize the bibliographic and demographic space (

Table **2**).

Table 1. Research questions

| RQ# | Research question | Description |
| --- | --- | --- |
| RQ1 | What project management approaches have been mentioned? | A list of project management approaches: agile, plan-based, etc. |
| RQ2 | What project management life cycles have been mentioned? | A list of project management lifecycle: waterfall, iterative, incremental, etc. |
| RQ3 | What characteristics of the project are mentioned to select each approach? | A list of characteristics: stable requirements, project size, etc. |
| RQ4 | What characteristics of the project are mentioned to select each life cycle model? | A list of characteristics: stable requirements, project size, etc. |
| RQ5 | What frameworks are mentioned to classify project management approaches? | A list of classification frameworks: Cynefin, Stacey matrix, etc. |
| RQ6 | What evidence is offered on results in real-world settings? | List of reported benefits or pending challenges. |

Table 2. Publication questions

| PQ# | Publication question | Description |
| --- | --- | --- |
| PQ1 | Where the studies had been published? | To know the distribution of studies by type of venue: conferences, journals, or workshops. |
| PQ2 | How has the quantity of studies evolved? | Publications per year. |

| PQ# | Publication question | Description |
|---|---|---|
| PQ3 | What are the authors' affiliations? | Classify the affiliations into two categories: academy or industry. We will consider the affiliations of all the authors. |
| PQ4 | Which are the most active countries? | Considering the author's affiliations (all authors). |

## 2.2. Search strategy and study selection

The selected search strategy includes two approaches to look for the primary studies. The first one is an automatic search on the most important online sources for scientific studies (digital libraries and databases). The second one, used in order to ensure the completeness of our set of studies, is the snowballing technique [9]. Figure 1 shows these strategies.

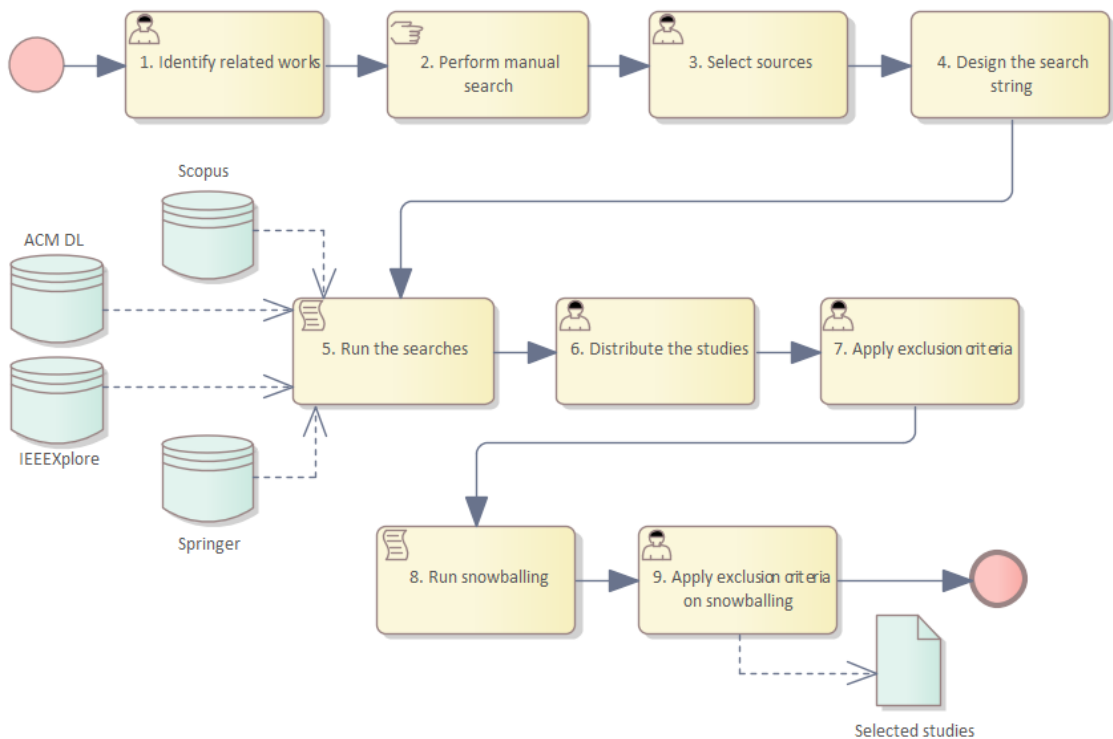

Figure 1. Search and selection process

The study selection strategy will include the classification and revision of every study in the set of retrieved works, aiming to select those relevant papers, regarding to the RQs, as presented in Figure 1 above.

The activities are as follows:

**Activity #1: Identify related works**
We will start our study identifying the related works previously conducted, as recommended by Petersen et al. [8], since this will also help us to adjust the focus of our study. As we are conducting a SMS, a type of secondary study, we will only consider as related work other secondary works (SMS or SLR) previously published.

**Activity #2: Perform manual search**
We have planned to conduct a manual of those related works, in order to calibrate our study.

**Activity #3: Select sources**
The electronic databases of scientific articles selected for this study are Scopus, IEEE Xplore, ACM digital library, and Springer, as they are cited repeatedly in SMS reports and guidelines [10] [11] [12] [13].

**Activity #4: Design the search string for each source**
The search terms chosen that will be run mainly on title, abstract and keywords belong to the categories stated by the PICO method (Petersen et al. [14]) as follows:

**Population:** we want to characterize project management approaches, that's why one of the selected terms is "project management".
**Intervention:** this category refers to software engineering areas, so we have selected the term "software".
**Comparison:** we want to compare agile and plan-based project management approaches, so we have selected the terms "(agile) AND (waterfall OR traditional OR classic* OR "plan-based")".
**Outcome:** we don't have settled terms for this category, but we'll look for the evidence indicated in the RQ (Table 1).

The specific search strings for each database are the following ones:

Table 3. Search strings

| Database | Search string |
| --- | --- |
| Scopus | TITLE-ABS-KEY((software) AND ("project management") AND (agile) AND (waterfall OR traditional OR classic* OR "plan-based")) |
| IEEE Xplore | (("Index Terms":software) AND ("Index Terms":"project management") AND ("Index Terms":agile) AND ("Index Terms": traditional OR "Index Terms":classic* OR "Index Terms":waterfall OR "Index Terms":"plan-based")) |

| Database | Search string |
|---|---|
| ACM DL | [All: software] AND [All: "project management"] AND [All: agile] AND [[All: waterfall] OR [All: traditional] OR [All: classic*] OR [All: "plan-based"]] |
| Springer Link | software AND "project management" AND agile AND (waterfall OR traditional OR classic OR "plan-based") |

**Activity #5: Run the searches**
The searches are executed, and the results collected. These results will contain duplicates that must be eliminated by applying these rules:
a. Expanded works (or expanded versions): keep the last one.
b. Duplicated works: depending on the source, following this priority order: Scopus (since it offers the most detailed information), followed by IEEE Xplore, then Springer, and, finally, ACM DL (because it does not retrieve the abstracts of the studies) [13].

**Activity #6: Distribute the studies**
The retrieved studies will be distributed among four researchers as Table 4 shows. Notice that we ensure that every single work will be examined by two different researchers, in order to reduce bias.

Table 4. Distribution of studies.

| Researcher | Studies | | | |
|---|---|---|---|---|
| | 0%-25% | 26%-50% | 51%-75% | 76%-100% |
| R1 | X | X | | |
| R2 | | X | X | |
| R3 | | | X | X |
| R4 | X | | | X |

The individual selection of studies made by each researcher will be consolidated into a unique set of studies. Differences among researchers will be solved by using the following criteria [8]:

Table 5. Criteria to resolve disagreements.

|  |  | Researcher 1 | | |
|---|---|---|---|---|
|  |  | Include | Uncertain | Exclude |
| Researcher 2 | Include | A | B | D |
|  | Uncertain | B | C | E |
|  | Exclude | D | E | F |

**A & B:** the study is included.
**E & F:** the study is excluded.
**C & D:** the paper is read in full and qualified again until obtaining A, B, E or F.

**Activity #7: Apply exclusion criteria**
The researchers will independently review the studies they have been assigned to and they will decide whether the studies are relevant or not, by only reading their title and abstract and then applying the exclusion criteria (EC). The set of retrieved studies will be then filtered by applying the exclusion criteria described below, in Table 6.

Table 6. Exclusion criteria.

| EC# | Description |
|---|---|
| EC1 | The study is not written in English. |
| EC2 | The study venue is not conference, workshop, or journal. |
| EC3 | The study is not peer-reviewed. |
| EC4 | Short papers (four pages or less). |
| EC5 | The focus is not on project management software development. |
| EC6 | The study doesn't discuss about the use of agile or plan-based project management approaches. |

**Activity #8: Run snowballing**
Resulting articles will be considered as "seed works" to be used on a forward and a backward snowballing technique, following the guidelines proposed by Wohlin [9]. The motivation for running a forward and backward snowballing complementary search aims to complement the automatic search and to collaborate with the search strings refinement.

SYSTEMATIC MAPPING PROTOCOL
AGILE STRATEGIES FOR SOFTWARE DEVELOPMENT ACCORDING TO TECHNICAL AND ENVIRONMENTAL
COMPLEXITY FACTORS## 2.3. Data extraction form

Relevant data are extracted from the set of selected studies to answer the eight RQs and the four PQs. Data are stored into a spreadsheet with the format shown in Table 7 (Data Extraction Form, DEF) and in Table 8.

Table 7. Data extraction form for RQ.

| Study #ID | RQ1 | RQ2 | RQ3 | RQ4 | RQ5 | RQ6 |
|---|---|---|---|---|---|---|
| Study #1 | | | | | | |
| Study #2 | | | | | | |
| … | … | … | … | … | … | … |
| Study #n | | | | | | |
| **Accepted values** | PM approach names (text) | PM life cycle model names (text) | Characteristic names (text) | Characteristic names (text) | Framework names (text) | Result names (text) |

We have selected different presentations depending on the number of possible results: when they may be a lot, we will use a bar chart, but we will employ a pie chart when could be a few.

Table 8. Data extraction form for PQ.

| Study #ID | PQ1 | PQ2 | PQ3 | PQ4 |
|---|---|---|---|---|
| Study #1 | | | | |
| Study #2 | | | | |
| … | … | … | … | … |
| Study #n | | | | |
| **Accepted values** | Fora names (text) | Year of publication (integer) | Academia Industry Research center | Country names (text) |

## 3. Threats to validity

In order to minimize the impact of the validity threats categorized by Petersen et al. [8] that could affect our study, we present them with the corresponding mitigation actions:

**Descriptive validity**

This validity seeks to ensure that observations are objectively and accurately described.

- We have structured the information to be collected by means of a couple of Data Extraction Forms, for RQs and PQs, presented in Table 7 and Table 8, to support an uniform recording of data and to objectify the data extraction process.
- Besides, all the researchers will participate on an initial meeting, aimed at unifying concepts and criteria, answer to any question and to demonstrate (by examples) how to conduct the data extraction process.
- We will also make public our data extraction form.

**Theoretical validity**

The theoretical validity depends on the ability to get the information that it is intended to capture.

- We will start with a search string (Table 3) tailored for the four most popular digital libraries on computer sciences and software engineering online databases.
- An expert will provide a set of articles to verify if they are retrieved with the search string.
- A set of exclusion criteria (Table 6) to objectivize the selection process have been defined.
- We will distribute the studies among four researchers, working independently and, with an overlap of studies that ensures that each study is reviewed by at least two researchers (Table 4).
- We will combine two different search methods: an automatic search and a snowballing, to diminish the risk of not finding all the available evidence.
- It could be a minimal impact due to the selection of articles written in English and the discard of other languages.

**Generalizability**

This validity is concerned with the ability to generalize the results to the whole domain.

- Our set of RQs is general enough in order to identify and classify the findings on aspect-oriented software development methodologies regardless specific cases, type of industry, etc. [14].

**Interpretive validity**

This validity is achieved when the conclusions are reasonable given the data.

- At least two researchers will validate every conclusion.
- Two researchers, experienced on the problem domain, will help us with the interpretation of data.

**Repeatability**

The research process must be detailed enough in order to ensure it can be exhaustively repeated.

- We have designed this protocol sufficiently detailed to allow to repeat the process we have followed.
- The protocol, as well as the results of the study, will be published online, so other researchers can replicate the process and, hopefully, corroborate the results.

# 4. Conclusions

We have strictly followed the guidelines published by Petersen et al. [8] to: plan, conduct and report a SMS. As the whole team adhered to these guidelines to build up the protocol presented in this document, we think the execution phase (conducting the SMS) will be rigorous, transparent, and repeatable, and the threats to validity have been mitigated as much as possible.